\documentclass[12pt]{iopjournal}
\usepackage{ragged2e}
\usepackage{upgreek}
\usepackage[style=numeric,sorting=none]{biblatex}
\usepackage{siunitx} 
\AtEveryBibitem{\clearfield{title}} 
\renewbibmacro{in:}{%
}

\begin{document}
\title{Towards Animate Droplets: Active, Adaptive, and Autonomous}

\author{Joe Forth$^{1,2,*}$, Robert Malinowski$^{3,*}$ and Giorgio Volpe$^{3,*}$}

\affil{$^1$Department of Physics, University of Liverpool, Crown Street, Liverpool L69 7ZD, United Kingdom}

\affil{$^2$Department of Chemistry, University of Liverpool, Oxford Street, Liverpool L69 7ZE, United Kingdom}

\affil{$^3$Department of Chemistry, University College London, 20 Gordon Street, London WC1H 0AJ, United Kingdom}

\affil{$^*$Author to whom any correspondence should be addressed.}

\email{j.forth@liverpool.ac.uk; robert.malinowski.15@ucl.ac.uk; g.volpe@ucl.ac.uk}


\begin{abstract}
\justifying
Droplets, sub-millilitre liquid volumes with at least one interface, have traditionally served as compartments for storing, transporting, and delivering materials. Beyond familiar applications in food, coatings, and consumer goods, they find cutting-edge use in energy storage, sensing, and tissue engineering. The next frontier is their integration into animate matter, emerging materials defined by their levels of activity, adaptiveness, and autonomy. Easy to produce and dispense or print into complex structures, and with enormous chemical versatility, droplets are ideal building blocks for animate matter. In this Perspective, we outline a roadmap for advancing animacy in droplets and call for a more concerted effort to integrate novel mechanisms for motility, sensing, and decision-making into droplet design. Although research on active droplets spans more than a century, achieving true autonomy, where droplets process multiple stimuli and respond without external control, remains a central challenge. We hope to inspire interdisciplinary collaboration towards applications in consumer goods, microfluidics, adaptive optics, tissue engineering, and soft robotics.
\end{abstract}

\justifying

\section*{Animacy in Droplets: The State of the Art}

Advances in materials science are placing a new class of synthetic materials to the fore: animate matter~\cite{volpe2024roadmap}. From protocells to living architectures, these systems are beginning to unsettle long-standing boundaries between inert matter and living systems, promising transformative impacts in the circular economy, health and climate resilience~\cite{volpe2024roadmap}. These materials are defined through three principles (the three A's of animacy), being active, adaptive and autonomous to various degrees~\cite{miodownik2021animate}. In this Perspective article, we argue that these same principles can and should be applied to guide the development of droplet research. These \emph{animate droplets} still consist of the \unit{\um}-to-mm-sized liquid compartments that are familiar from everyday life, but are modified to incorporate mechanisms to convert energy from and respond to their surroundings. In general, the simplest animate droplets can be made from a three-component system of oil, water, and a surfactant, in which gradients in surface tension generate motion, shape change, or division~\cite{ray2023simple}. Animate droplet systems can vary in complexity in terms of composition, structure, and population. Chemical reaction networks can be easily compartmentalised in small liquid volumes, with reagent exchange mediated both through the surrounding liquid and via direct connections between droplets such as molecular pores~\cite{delgado2011coupled,villar2011formation}. Provided the individual droplets are stabilised against coalescence, collective behaviours can emerge where exchange of chemical information causes dynamical clusters and patterns to emerge that are reminiscent of those seen in complex biological systems~\cite{hokmabad2022chemotactic}. The structural complexity of individual droplets can be increased by introducing asymmetry, droplets-in-droplets, or chirality to the droplet or its surrounding phase~\cite{kruger2016curling,jin2017chemotaxis,dwivedi2021rheotaxis}, leading to spiralling motion and precession~\cite{hokmabad2019topological,wang2021active}. 

\begin{figure}[!h]
\includegraphics[width=\linewidth]{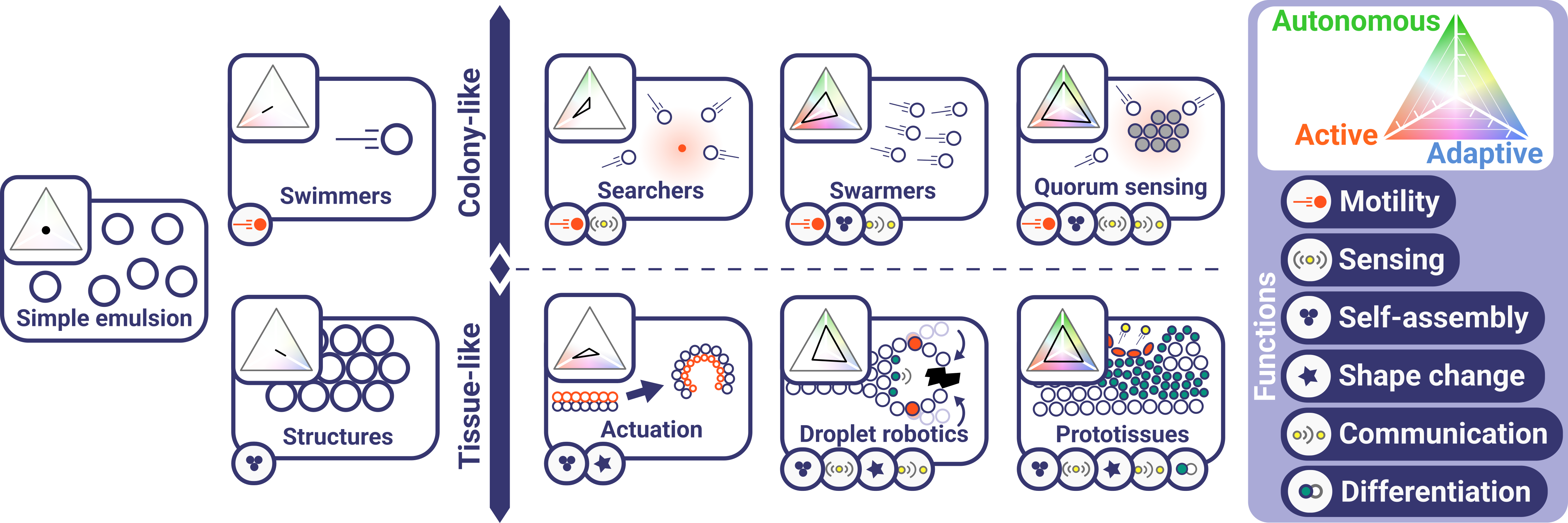}
\caption{\textbf{Animacy in droplets: the state-of-the-art.} To date, droplet systems have been designed to express behaviours embodying activity, adaptiveness, and autonomy to various degrees. These behaviours can achieve various elementary functions such as motility, sensing, self-assembly, shape change, communication, and differentiation (legend symbols). Starting from a simple emulsion, where droplets have stability against coalescence, combinations of these functions (inset symbols) can be used to achieve higher-order behaviours, such as swimming (swimmers) and self- or directed assembly (structures). As these traits are combined and the emergent complexity of the behaviour increases, so in general does the extent to which the system exhibits activity, adaptiveness, and autonomy (inset spider plots). Each behaviour can be considered more ``colony-like" (top row) or ``tissue-like" (bottom row) depending on whether droplets tend to act more like individual units working together or as mechanically bound pieces in a greater structure.}
\label{fig_taxonomy}
\end{figure}

In numbers, these systems range from single droplets to thousands or more. With increased numbers comes the ability to vary how droplets couple to one another, which in turn leads to different classes of collective behaviour (Fig.~\ref{fig_taxonomy}). Reversible interactions between droplets with weak mechanical coupling may occur through hydrodynamics~\cite{liu2019reconfigurable}, exchange of chemical reagents~\cite{krishna2024dynamic}, and capillary interactions~\cite{thapa2025capillary,karpitschka2016liquid}. In such systems, droplets behave as individual agents in which exchange of information leads to ensembles exhibiting \emph{colony-like} behaviour including `searching' for a chemical source by swimming up a concentration gradient~\cite{jin2018chemotactic}, swarming~\cite{thutupalli2011swarming}, and quorum sensing~\cite{de2024quorum} (Fig. \ref{fig_taxonomy}, top row). Alternatively, additive manufacturing or directed assembly methods can be used to make mechanically coupled ensembles that are joined through surface forces~\cite{alcinesio2020controlled} or chemical bonding~\cite{mcmullen2022self} (Fig. \ref{fig_taxonomy}, bottom row). These \emph{tissue-like} constructs can incorporate structural heterogeneities such as variations in salt concentration that, when droplets are selectively connected, lead to structured mass transfer within the printed structures and hence actuation~\cite{villar2013tissue}. Stimulus-responsive components such as magnetic particles can, when incorporated into complex printed droplet structures, produce `droplet robotic' devices that perform useful tasks like dragging material through a maze~\cite{zhu2022aquabots}. Combining heterogeneous droplet composition and stimulus response leads to printed liquid systems with regional specialisation, so-called droplet \emph{prototissues}. So far, such prototissues have been produced with the ability to fabricate their own spatially specific transportation pathways via the synthesis of transmembrane pores~\cite{booth2017light} and to exhibit adaptive restructuring in response to electrical stimulus~\cite{fica20253d}. As with animate matter more generally, animate droplets can be discussed according to three recently articulated principles (the three A's) of animacy \cite{volpe2024roadmap}: activity, adaptiveness, and autonomy, with the latter being the least explored in droplets to date.

\textbf{Activity} implies the ability to convert energy to perform tasks and work~\cite{volpe2024roadmap}. Motility through self-propulsion is arguably the most intuitive example of activity. Motile droplets, like other active matter~\cite{bechinger2016active}, break symmetry to self-propel, most often because of an asymmetric surface tension caused by a temperature or compositional gradient \cite{malinowski2020advances, malinowski2020nonmonotonic}. Activity can also lead to shape change, growth, and division~\cite{armstrong2013butschli}. In droplets, this is often caused by the presence of some combination of surfactants that drastically alter surface tension~\cite{zhu2009coupled}. These shape changes may also couple to motion; such an example exists in the form of hydrocarbon droplets in an aqueous solution of low-molecular-weight surfactant. As the system cools, the surfactant nucleates a liquid crystalline microphase at the droplet surface~\cite{denkov2015self}. These phase transition leads to a rich range of droplet shapes including facets, doughnuts, and tendrils, while oscillating the temperature of the system leads to reversible shape changes that drive motion~\cite{cholakova2021rechargeable}. Technically far more challenging is the incorporation of an active cytoskeleton within a droplet. Magnetic particles can form active cytoskeletons when driven by an external field, producing droplets that move, change shape, and divide; however, their reliance on an external field limits their autonomy~\cite{kim2025shape}. More directly biological are motile assemblies of tubulin fibrils and kinesin molecular motors assembled at the oil-water interface that drive droplet motion~\cite{sanchez2012spontaneous}.

\textbf{Adaptiveness} is the ability to individually and collectively switch between dynamical states by sensing and responding to internal dynamics and environmental changes~\cite{volpe2024roadmap}. Rudimentary adaptive behaviour can be introduced into droplets by adding a light-responsive component to an active system. For example, molecular motors with light-switchable chirality produce spiralling motion in droplets whose swimming handedness is controlled by light~\cite{lancia2019reorientation}. A degree of adaptiveness has also been introduced into droplet clusters that had been mechanically coupled via biotin-streptavidin bonds. These droplet oligomers could sense single droplets through the exchange of solutes, resulting in chasing behaviour~\cite{kumar2024emergent}. To date, the most compelling demonstrations of adaptive droplets are systems composed of two active species of differing chemical composition that exchange reagents asymmetrically, leading to switching between dynamical states dominated by either individual or collective droplet behaviours~\cite{birrer2022we}. These effects are seen most commonly in Marangoni swimmers that exchange surface-tension-modifying chemicals asymmetrically, leading to non-reciprocal interactions that, at a coarse-grained level, violate Newton's third law~\cite{meredith2020predator}.

\textbf{Autonomy} is the ability to process multiple information inputs and choose between a diverse repertoire of responses without external intervention~\cite{volpe2024roadmap}. Arguably, even simple Marangoni swimmers exhibit rudiments of autonomy, as they can `sense' their environment and `decide' their direction of motion, for example, when droplets navigate a maze successfully~\cite{lagzi2010maze}. However, it could also be easily argued that these droplets simply follow a concentration gradient under topographical constraint. More compelling examples of autonomy are droplets that incorporate chemical oscillators, most often Belousov-Zhabotinsky (BZ) reactions~\cite{vanag2001pattern,duenas2019chemistry}. Encapsulating the BZ reaction within droplets enables the decoupling of transport time- and length-scales for polar excitatory species (e.g., bromous acid) and non-polar inhibitory species (e.g., bromine). This separation gives rise to rule-based, time-evolving patterns of droplet ensembles that emerge in response to defined chemical inputs~\cite{tompkins2014testing}. Theoretical work has argued such systems are capable of solving computational problems such as Boolean satisfiability~\cite{guo2021molecular}, suggesting the capability for complex decision-making in droplet ensembles. However, experimental realisations of such a system are currently lacking. A shortcut for engineering autonomous droplet systems is to directly incorporate inherently autonomous biological components such as living cells~\cite{vladescu2014filling,xu2022living,graham2017high}. However,
achieving true synthetic autonomy, where droplets can sense, decide, and act without external input, remains a major frontier. 

\section*{Future Directions - Inspiration from Biology and Beyond}
Inspiration for making droplets that exhibit high degrees of animacy, especially less-explored traits such as autonomy, can come from both nature and affine fields like active matter and soft robotics ~\cite{bechinger2016active,kriegman2020scalable}. Drawing on ideas from both living and synthetic systems, we briefly outline some particularly promising avenues to engineer animacy in droplets (Fig. \ref{fig:2}).

\subsection*{Sustained Energy} 
\begin{figure}
\includegraphics[width=\linewidth]{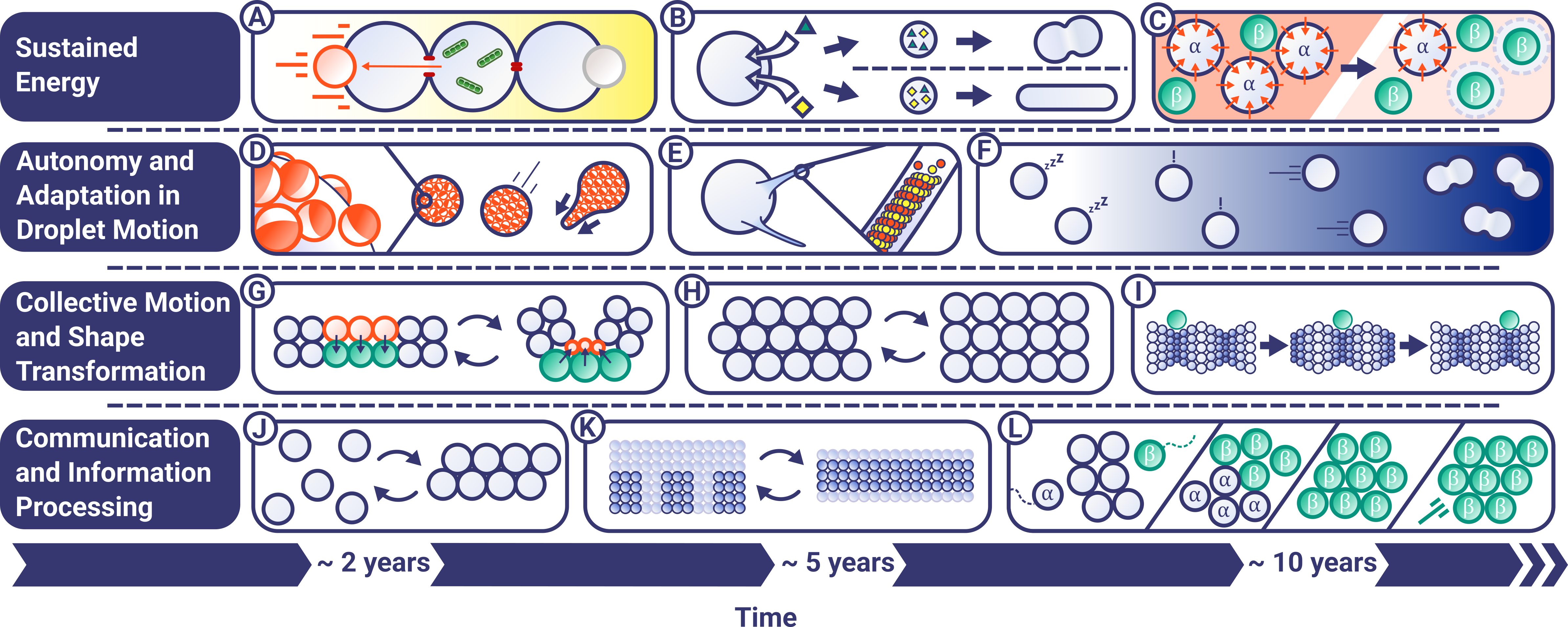}
\caption{{\bf A roadmap for research in animate droplets}. Our suggestions, left to right, start from simpler advances on current literature to abstract towards more complex future implementations.
{\bf Sustained Energy.} 
{\bf A}: Incorporation of chloroplasts (green) and active particle propulsion into a multi-droplet structure connected by pores could create a  liquid machine sensitive to light (yellow colouring).
{\bf B}: Droplets that produce different responses, such as division or shape change in response to different concentrations of chemical stimuli (triangles and diamonds) 
{\bf C}: Complex droplet systems that exhibit population control, switching high energy demand units ($\upalpha$) for lower energy demand units ($\upbeta$) when nutrient sources become scarce (orange colouring). 
{\bf Autonomy and Adaptation in Droplet Motion:} 
{\bf D}: Active particle Pickering emulsions allowing motion and shape change.
{\bf E}: Internal control of a biomimetic droplet cytoskeleton for autonomous deformation.
{\bf F}: Adaptive droplets that seek nutrient sources to grow and divide only when favourable.
{\bf Collective Motion and Shape Transformation:}  
{\bf G}: Cyclical actuation in droplet prototissues driven by reversible active pumping of solutes.
{\bf H}: Structural changes, such as moving from a hexagonal to a cubic pattern could be implemented by changes in the nature of inter-droplet adhesion, as opposed to relying on specialised actuators, as in (G).
{\bf I}: Cilia-like transport of cargoes driven by controlled waves of droplet contraction in synchronised chemical oscillators.
{\bf Communication and Information Processing:} 
{\bf J}: Droplet structures reversibly switching between colony-like and tissue-like behaviour by varying the strength of mechanical coupling in the system. 
{\bf K}: Plastic self-rearrangement of prototissue structure, allowing a change in subunit distribution and/or prototissue function. For example, separate compartments (dark blue) in a supporting matrix (light blue) could be reorganised into a connected channel to allow transport of materials.
{\bf L}: Information exchange leading to a change in droplet phenotype (from $\upalpha$ to $\upbeta$) and collective decision-making (from quiescent to active) through droplet quorum sensing. 
}
\label{fig:2}
\end{figure}

Autonomous droplets must power their own processes with energy cycles similar to those in living cells. Due to the ease by which it can be generated, manipulated, and controlled, light is an excellent candidate energy source for progress here~\cite{rey2023light}. Synthetic droplet systems capable of autonomous generation of ATP and NADPH upon light illumination have already been produced through the encapsulation of chloroplasts~\cite{miller2020light}. Harnessing this system to produce shape change or motion could readily be achieved in droplets connected by pores~\cite{elani2014vesicle}, where reaction products from chloroplast-containing droplets diffuse into compartments that are surface-functionalised with motility-driving components such as enzymes and particles (Fig. 2A). Non-biological energy cycles also abound in the literature due to the rapid growth of systems chemistry as a formal discipline~\cite{ashkenasy2017systems}. Dissipative coacervates that undergo phase separation into droplets-within-droplets~\cite{sastre2025size} are a fascinating example of a chemical oscillator that couples to structure. To advance these dissipative droplets towards autonomy, the chemical reaction cycles driving phase separation must incorporate a palette of dynamical states to produce droplets that phase-separate or divide on-demand in response to complex chemical conditions being satisfied (Fig. \ref{fig:2}B)~\cite{maguire2019importance}. A more distant goal is self-regulation, where animate droplets process external stimuli (e.g., local concentration of chemical fuel) and adapt their behaviour accordingly, either accelerating or slowing chemical activity (Fig. \ref{fig:2}C). Designers of self-sustaining animate droplet ensembles must draw on ecosystem modelling~\cite{may2001stability}, dynamical cellular process regulation~\cite{kholodenko2006cell}, and control theory~\cite{alvarado2025optimal} to design ecologies of animate droplets that can regulate their population, growth rate, or internal processes to avoid resource exhaustion.

\subsection*{Autonomy and Adaptation in Droplet Motion}
Swimmers that use Marangoni stresses to move are the most promising systems for introducing decision making into motile droplets; the underlying physics describing their motion is fairly well understood and a wide range of chemistries can be used to make them~\cite{maass2024self}. Propulsion mechanisms beyond Marangoni stresses must also be explored. This could be achieved by adding active particles into Pickering emulsions (Fig. \ref{fig:2}D)~\cite{ding2016microfluidic} or, more challenging, by extending existing work on active droplets incorporating biological cytoskeletal components to produce shape-changes or crawling (Fig. \ref{fig:2}E). 
More speculative is the fabrication of droplets that can exploit the ability of chemical oscillators to switch between dynamical states in response to external triggers~\cite{ter2023catalytically}, such as searching for chemical fuel in regions with low fuel concentration and growing or dividing in regions with high fuel concentration (Fig. \ref{fig:2}F). The dissipative coacervate droplets discussed above offer a promising avenue for achieving this~\cite{donau2020active,poprawa2024active}: Autonomy could be increased by linking the synthesis of degradable, motion-driving surfactant-like molecules to the chemical reaction networks that govern droplet phase separation. The inklings of this auto-chemotactic behaviour have already been reported in a system of water droplets dispersed in oil, in which each droplet incorporates a lipid precursor that reacts with a second component in the oil phase to form micelles~\cite{babu2021acceleration}, driving motion towards regions rich in precursor compound. 
    
\subsection*{Collective Motion and Shape Transformation}
Despite the compositional complexity that can be produced in printed droplet prototissues, energy conversion mechanisms have rarely been incorporated into these systems to produce autonomous, or even active, prototissues. Achieving this goal could be as simple as integrating existing active droplet designs with one of the many available open-source inkjet printing or droplet dispensing methods~\cite{li2018controllable}. This approach could produce energy converting prototissues that undergo cyclical or stimulus-driven actuation (Fig. \ref{fig:2}G), rather than simply relaxing to equilibrium as has been achieved to-date. Varying the strength of mechanical coupling between animate droplets is likely to lead to novel collective modes of actuation, ranging from chiral waves similar to those observed in precessing starfish embryos~\cite{tan2022odd} to the ability to switch between colony-like and tissue-like states in order to undergo complex structural reconfiguration based on processing surrounding chemical conditions (Fig. \ref{fig:2}H). In the case of strong mechanical coupling in prototissues, mechanical and chemical stimuli in droplets have the potential to lead to orikata-like folding from two-dimensional templates or adaptive cilia-like wave actuation to achieve directed mass transport (Fig. \ref{fig:2}I)~\cite{zhou2019biasing}. None of these phenomena have yet been observed in droplet-based systems.
    
\subsection*{Communication and Information Processing} 
Truly autonomous systems can process multiple inputs to decide on a course of action. Most striking about droplet systems is their ability to exist in both colony-like and tissue-like states depending on the strength of mechanical coupling between the droplets, and hence exploit the fundamentally different rules by which these systems process information~\cite{sole2019liquid}. Colony-like responses are often mediated by indirect communication through the environment \cite{dias2023environmental}, as seen in the quorum sensing of bacteria and the stigmergy of social insects~\cite{sinclair2022model}. By contrast, tissue-like systems process information by exchange of information along persistent architectures. The possibility of dynamically tuning the mechanical coupling between droplets potentially allows droplet systems to switch between these two states (Fig. \ref{fig:2}J). Communication similar to quorum sensing has been replicated in groups of droplets relaxing to equilibrium~\cite{de2024quorum}. The next step would be to implement these processes in an energy-converting droplet system. Incorporation of light-sensitive BZ reactions or other stimulus responsive chemical oscillators into droplet prototissues, combined with emerging paradigms describing the performance of computational operations by such systems~\cite{guo2021molecular}, could provide a powerful foundation for systems that use energy conversion to switch between different states, e.g., dynamical patterns. The use of such systems is not just in solving computational problems, but also in designing structures that are either self-building or switch between different structures autonomously (Fig. \ref{fig:2}K). Moving away from traditional computational descriptions of information processing, there exist a number of recent theoretical predictions of state change through quorum sensing phenomena in droplet systems that exhibit simple information exchange and non-reciprocal interactions~\cite{ziepke2022multi,duan2023dynamical} (Fig. 2L). The chemical frameworks for producing these behaviours in droplets already exist, harnessing them to produce truly autonomous droplet communities within a decade seems an eminently feasible goal.

\section*{Outlook - Towards True Animate Droplets}
Achieving fully animate droplets capable of purposeful adaptive and autonomous behaviours will require integrating mechanisms for energy harvesting and conversion, alongside sophisticated control over actuation, sensing, memory, communication, and information processing. This requires the adoption of an inherently interdisciplinary approach, blending novel fabrication techniques such as digital microfluidics and multi-component ink-jet printing with emerging paradigms in chemical oscillators and chemical reaction networks. These collaborative approaches are crucial for uncovering new physics, including non-reciprocal interactions \cite{fruchart2023odd} and dynamical pattern formation in the presence of information exchange between autonomous agents \cite{saha2025effervescence}, phenomena not captured by classical equilibrium theories. Animate droplets could therefore be powerful experimental platforms for probing the fundamental principles of life-like behaviour in non-living matter and unlock transformative applications in smart drug delivery, microfluidics, adaptive optics, tissue engineering, and soft robotics as well as more every day applications like consumer goods. 

\subsection*{Acknowledgments} 
RM and GV gratefully acknowledge the Engineering and Physical Sciences Research Council for supporting this work [grant numbers EP/W005875/1, EP/W026813/1].

\printbibliography
\end{document}